\documentclass[final,5p,times,
numbers]{elsarticle}

\usepackage{amssymb}
\usepackage{caption}
\usepackage{array}
\usepackage{graphicx} 
\usepackage{amsthm}
\usepackage{lineno} 
\usepackage{adjustbox}
\usepackage{multicol} 
\usepackage{multirow} 
\usepackage{amsmath}
\usepackage{amsfonts}
\usepackage{graphicx}
\usepackage{dcolumn}
\usepackage{hyperref}

\journal{Nuclear Instruments and Methods in Physics Research A}

\usepackage{upgreek}
\usepackage{placeins}
\usepackage{xcolor}

\DeclareCaptionFormat{figures}{\textbf{#1#2} {#3}}
\DeclareCaptionFormat{tables}{\textbf{#1#2} {#3}}
\usepackage{microtype} 
\hypersetup{
   colorlinks=true
}
\makeatletter\def\Hy@Warning#1{}\makeatother

\begin{document}
\definecolor{ElsevierBlueish}{RGB}{0, 128, 172}
\hypersetup{
    linkcolor=ElsevierBlueish,
    citecolor=ElsevierBlueish,
    urlcolor=ElsevierBlueish
}
\hbadness=10001

\begin{frontmatter}



\title{Design, development, and construction of the new beam stoppers for CERN's injector complex}


\author[a]{D.~Baillard\corref{cor1}}
\ead{dylan.baillard@cern.ch}
\author[a]{E.~Grenier-Boley}
\author{M.~Dole}
\author{F.~Deslande}
\author[a]{R.~Froeschl}
\author[a]{T.~Lorenzon}
\author[a]{P.~Moyret}
\author[a]{R.~Peron}
\author{A.~Pilan~Zanoni}
\author[a]{C.~Sharp}
\author[a]{M.~Timmins}
\author[a]{and M.~Calviani\corref{cor1}}
\ead{marco.calviani@cern.ch}

\affiliation[a]{organization={CERN},
            addressline={Esplanade des Particules, 1},
            city={Genève},
            postcode={1211},
            country={Switzerland}}

\cortext[cor1]{Corresponding authors}

\begin{abstract}
Beam stoppers are installed in the transfer lines of the CERN accelerator complex; these components are used as part of the access safety system, which guarantees the safety of workers in the accelerators. They are designed to stop one or at most a few pulses of the beam, where ``stop'' means the partial or complete absorption of the primary beam in such a way that the remaining unabsorbed primary or secondary beam remains below a specified threshold, as defined by the needs of radiation protection. Prior to Long Shutdown 2 (LS2; 2018--2021), beam stoppers in the injector complex were dimensioned for beam-pulse energies between 9.0 and 30~kJ. The upgrade of the accelerator complex in the framework of the LHC Injectors Upgrade (LIU) project involves beam-pulse energies of up to 92.5~kJ, meaning that these beam stoppers are not able to fulfill the new functional specifications. To cope with the LIU beam parameters and fulfil requirements for safety, maintainability, efficiency, and reliability, a new generation of 28 beam stoppers has been designed, built, and installed. The aim of this paper is to demonstrate the requirements-driven design of these new beam stoppers, outlining the main requirements along with a description of the design and structural assessments. This document presents the implementation and integration of a standardized but adaptable design using a unique 564-mm-long stopper core with a CuCr1Zr absorber and an Inconel~718 diluter, taking into account radiological and infrastructure challenges. The installation process is also described, and the first operational feedback received since LS2 is presented.
\end{abstract}




\end{frontmatter}


\section{Introduction}
\label{Introduction}
CERN's Proton Synchrotron (PS) beam can be sent to several different accelerators and experimental facilities. The PS has been in operation since 1959 and has undergone various upgrades for physics-performance and safety reasons. Among these upgrades, the installation of the first beam stoppers took place in the 1970s. Beam stoppers consist of an absorbing stopper core that remains out of the ultra-high-vacuum particle beam line (in the ``out-beam'' position) during normal operation of the accelerators \cite{Maglioni}. In normal conditions, to give access to downstream facilities, the core (the absorbing material) of a beam stopper is moved into the beam line (the ``in-beam'' position) by an operator once the beam has been dumped upstream, so the stopper does not intercept the beam. In emergency scenarios, a beam stopper is moved automatically (by the access safety system) into the in-beam (``safe'') position to potentially intercept the beam. This occurs, for example, when someone forces an access door in a downstream accelerator and before the beam interlock can intervene to dump or deviate the beam upstream, to limit the exposure of personnel to ionizing radiation.


As part of the LHC Injectors Upgrade \cite{Damerau:1976692}---which aims to increase the maximum proton beam-pulse energy to 92.5~kJ ---an\-a\-lyt\-i\-cal studies~\cite{Pilan} have demonstrated that the existing beam stoppers will be inadequate for the new beam power. These are currently equipped with stainless-steel stopper cores that were originally designed for a beam-pulse energy of just 9.0~kJ, and they reach their structural limits after only one short pulse at this energy level. According to the limit allowed by the access safety system, the beam stoppers will need to withstand to at least five repeated pulses. As such, they are no longer able to fulfil their function under the upgraded beam scenarios in operation after Long Shutdown 2 (LS2)(2018-2021) of the LHC. Moreover, scarce documentation and the existence of several different designs make maintenance of the current beam stoppers difficult, supporting the need for the construction of new beam stoppers.

This paper describes the justification for the chosen diluter--absorber stopper-core configuration in~\cite{Pilan}, which represents a compromise between the desire to have a compact design (hence the selection of material with a lower nuclear inelastic scattering length) and reduced radioactivation properties. Secondly, it describes the demonstrator prototype that was required to validate the working principle with respect to the safety rules for personnel-protection devices.

The purpose of this document is to present a step-by-step definition of the needs and constraints of this equipment in CERN the injector complex, showing how the chosen design meets these needs using structural simulations. It highlights the challenges and advantages of using a standard and adaptable design for integration with different beam parameters and at different positions in the injector complex while taking into account radiation-protection considerations. This paper also presents the installation sequence of the beam stoppers, including removal, assembly, and installation steps. It concludes by presenting operational and mechanical-performance data obtained after 3~years of operation.

\subsection{Benchmark of the actual beam stoppers at CERN}
\label{Actual beam stoppers}
Up to 2019, there were more than ten different beam-stopper designs in the accelerator complex, including six designs in the injector lines. These are a result of a long evolution over time and specific needs in the different areas of the accelerator complex.

\begin{figure}[htb]
\begingroup\centering
\includegraphics[scale=0.12]{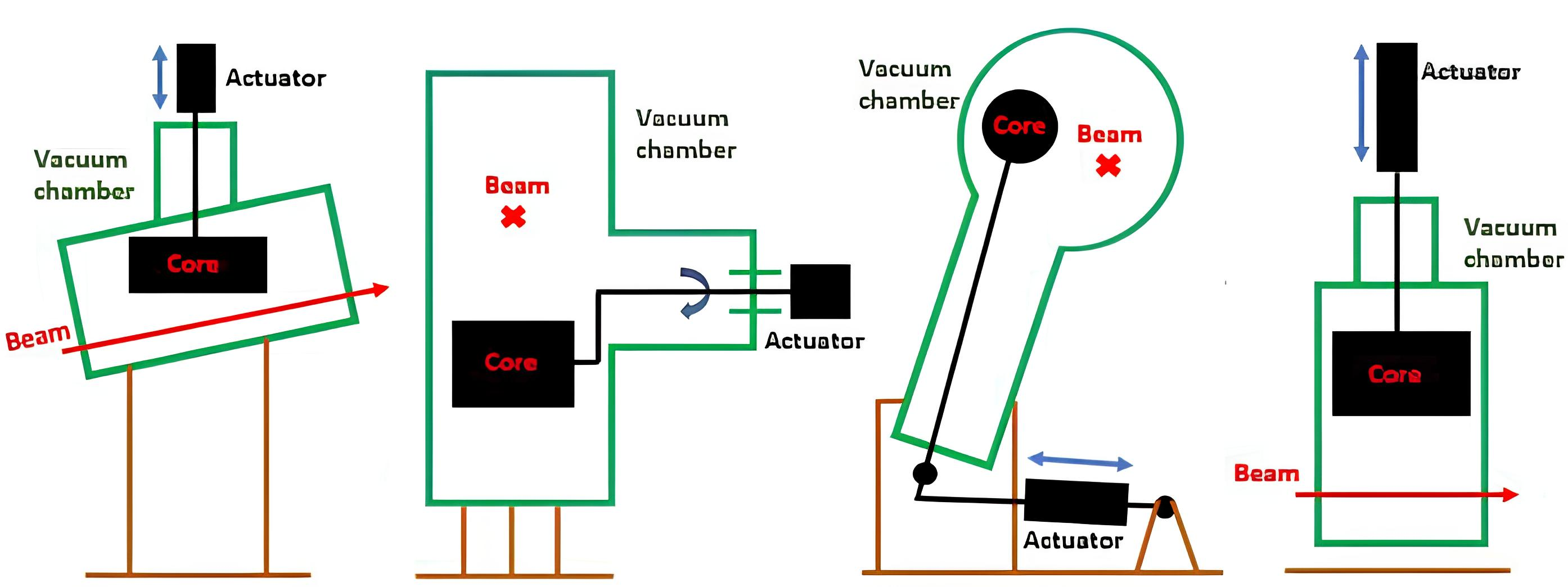}
\caption{Four examples ((a)--(d) from left to right) showing simplified designs of different beam stoppers installed in CERN's injector complex. Adapted from~\cite{Pilan}.}
\label{Bemact}
\endgroup
\end{figure}

The working principle of the design shown in Fig.~\ref{Bemact}(b) is a pivot mechanism in which the beam impacts the stopper core radially. In contrast, the other beam stoppers (Figs.~\ref{Bemact}(a), \ref{Bemact}(c), and \ref{Bemact}(d)) are composed of stainless-steel or iron cylinders of different lengths that are impacted axially and moved by a linear actuator. The stopper in Fig.~\ref{Bemact}(c) is composed of one or two stopper cores in a vacuum chamber, which are moved simultaneously by the control system in approximately 17~s. The stopper in Fig.~\ref{Bemact}(a) uses a linear movement to close the aperture in 4~s, whereas the stopper in Fig.~\ref{Bemact}(d) closes the aperture in 30~s using a pneumatic actuator.

\begin{table}[b]
\caption{Beam parameters according to the LHC Intensity Upgrade project at the PS extraction.~\cite{Pilan}}
\begin{tabular}{ p{3.5cm}>{\centering\arraybackslash}p{3.5cm}}
Particle type & Protons \\
 \hline
Beam momentum & 26~GeV/$c$ \\
Intensity per pulse & $2.3 \times 10^{13}$ particles \\
Pulse energy & 92.5~kJ \\
Total pulse length & 1.1~$\upmu$s \\
Beam size ($\sigma_{\mathrm{h}} / \sigma_{\mathrm{v}}$) & 1.50~mm / 1.39~mm \\
 \hline
\end{tabular}
\label{Beamparam}
\end{table}

\begin{table*}[htb]
\caption{Chemical composition, heat-treatment parameters, and mechanical properties of Inconel~718 diluter according to ASTM~E8.\cite{ASTM}}
\footnotesize
\begin{tabular} {m{3.1cm}|>{\centering\arraybackslash}m{1.4cm} > {\centering\arraybackslash}m{1.4cm} > {\centering\arraybackslash}m{1.4cm} > {\centering\arraybackslash}m{1.4cm} > {\centering\arraybackslash}m{1.4cm} > {\centering\arraybackslash}m{1.4cm} > {\centering\arraybackslash}m{1.4cm} > {\centering\arraybackslash}m{1.4cm}}
\hline\hline
\multirow{4}{3.1cm}{Chemical composition [wt\%]} & C & Si & Mn & P & Cr & Mo & Ni & Cu\\
& 0.022 & 0.08 & 0.08 & 0.008 & 17.85 & 2.97 & 53.7 & 0.05\\
& Co & Ti & Al & Nb & B & Fe & N &\\
& 0.34 & 0.95 & 0.53 & 5.23 & 0.0028 & 18.10 & 0.005\\
\hline
\multirow{2}{3.1cm}{Heat treatments} & Temp. & Soak & \multicolumn{3}{c}{Furnace cooling} & Temp. & Soak & Cool. \\
~ & $718^{\circ}$C & 8~h & \multicolumn{3}{c}{$\leq$37.8$^{\circ}$C/h} & $621^{\circ}$C & 8~h & Air \\
\hline
\multirow{3}{3.1cm}{Mechanical properties} & & YS [MPa] & \multicolumn{2}{c}{UYS [MPa]} & A [\%] & HBW & \multicolumn{2}{c}{Density [g\,cm$^{-1}$]} \\
& Ref. & $\geq$150 & \multicolumn{2}{c}{$\geq$185} & $\geq$12 & $\geq$331 & \multicolumn{2}{c}{\multirow{2}{*}{8.19}} \\
& Act. & 157.8 & \multicolumn{2}{c}{192.6} & 19 & 400\\
\hline\hline
\end{tabular}
\label{Incoprop}
\end{table*}

\begin{table*}[htb]
\caption{CuCr1Zr absorber designation, chemical composition, and mechanical properties according to EN~12420. \cite{EN12420}}
\footnotesize
\begin{tabular}{m{1.6cm}>
{\centering\arraybackslash}m{1.6cm} > {\centering\arraybackslash}m{2.1cm} > {\centering\arraybackslash}m{0.9cm} > {\centering\arraybackslash}m{0.9cm} > {\centering\arraybackslash}m{1.2cm} > {\centering\arraybackslash}m{1.2cm} > {\centering\arraybackslash}m{1cm} > {\centering\arraybackslash}m{1cm} > {\centering\arraybackslash}m{1cm} > {\centering\arraybackslash}m{1cm} }
 \hline\hline
Symbol & Number & Standard & Density [g\,cm$^{-1}$] & Cu & Cr [wt\%] & Zr [wt\%] & Rp0.2 [MPa]  & Rm [MPa] & A5 [\%] & HB \\
\hline
\multirow{2}{1.5cm}{CuCr1Zr} & \multirow{2}{1.5cm}{CW106C} & \multirow{2}{1.9cm}{EN 12420:1999} & \multirow{2}{1.5cm}{8.90} & Ref. & 0.5--1.2 & 0.03--0.30 & $\geq$270 & $\geq$360 & $\geq$15 & $\geq$110\\
& & & & Act. & 0.92 & 0.11 & 271 & 409 & 29.1 & 122\\
\hline\hline
\end{tabular}
\label{Cuprop}
\end{table*}

\section{Methodology}
\label{Main technical requirements}
As the core is the main component of a beam stopper, the first step was to determine its design~\cite{Pilan}. The methodology used to validate the beam-stopper design was to determine its structure and beam-attenuation performance based on safety requirements while considering mechanical and integration constraints in parallel. This allowed the stopper core to be designed and a demonstrator to be constructed highlighting the mechanical forces exerted and validating the mechanical behavior before it went into production. This validation of the operating principle allowed the continuation of the design, assessment, and production of the other sub-assemblies of all the beam stoppers.

\subsection{Standard stopper-core design}
\label{Standard stopper core design determination}
Taking the longest stroke time from the out-beam position to the in-beam position of the present beam stoppers, which is 17~s, and the shortest beam-repetition rate in the PS, which is 1.2~s, a stopper core could theoretically be impacted by 15 repeated beam pulses. This number is defined as the thermo-mechanical requirement and is coupled to the beam parameters~\cite{Pilan}. Even though the PS access safety system will only allow a maximum of five repeated pulses, 15 repeated pulses is considered as a conservative reference for the design of a thermomechanically robust stopper core, anticipating possible future beam upgrades. Furthermore, this approach is additionally conservative because each beam pulse would hit different positions on the stopper core while it was traveling between the out-beam and in-beam positions.

Due to the low likelihood of a beam impact in the stopper core, no significant material activation is expected for the stopper. Interaction of the beam and its secondary particles with the stopper cores is necessary to achieve the desired attenuation. However, this will inevitably create stray radiation.

Ref.~\cite{Pilan} demonstrated that single-block configurations made from several different types of material would not be able to withstand a proton beam with the parameters described in Table~\ref{Beamparam} while staying within the yield-strength limit. Moreover, the requirement to limit the temperature rise while having an integration constraint of an 884-mm total length for installation in the accelerator is paramount, allowing a modular installation in the PS lines. Thus, the choice of the material for the stopper core is a compromise between the material's activation properties and a short nuclear interaction length. A shorter hadronic interaction length will mean that the total length of the beam-stopper system can be shorter~\cite{Pilan}, which is a major advantage for its integration into existing facilities in terms of equipment length. In this respect, a material of very high density, i.e., a high-Z material in general, would be optimal. The chosen materials are the best compromise between all these requirements.

The primary beam attenuation is defined by:
\begin{equation}
    I=I_{0}e^{-\frac{L}{\lambda}},
\end{equation}
where: $I$ is the intensity of the primary beam, which does not undergo any inelastic scattering in the stopper core material; $I_{0}$ is the primary beam intensity impacting the stopper core; $L$ is the total length of the stopper core; and $\lambda$ is the nuclear interaction length. The required attenuation is between $3.9\lambda$ and $19.2\lambda$ depending on the required attenuation ~\cite{Pilan}. It has been specified that the actual beam attenuation shall be maintained, along with the 200-mm stopper-core diameter~\cite{Pilan}.

\begin{figure}[htb]
\centering
\includegraphics[scale=0.25]{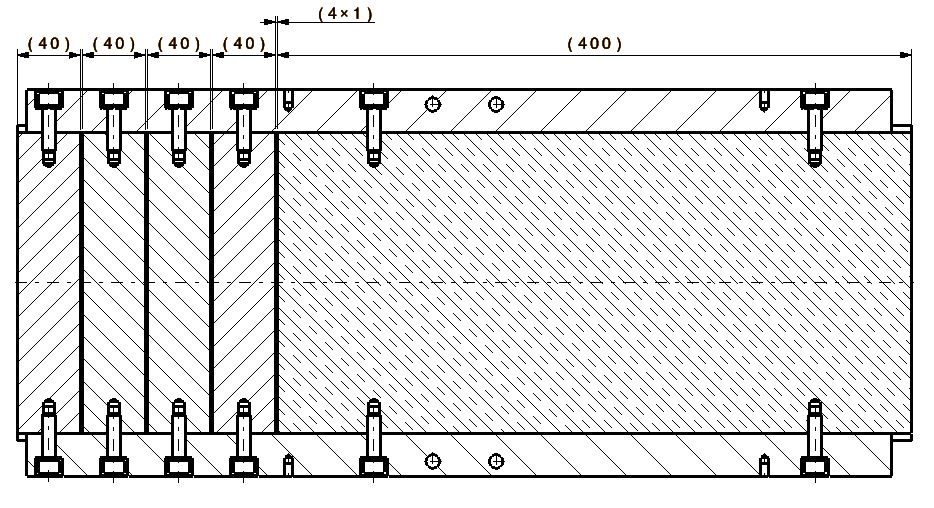}
\caption{Section view of the stopper-core, showing the four 40-mm-thick slices of Inconel~718 diluter spaced by 1~mm and the 400~mm CuCr1Zr absorber assembled together, dimensions in mm, beam coming from the left side of the picture. Design adopted from~\cite{Pilan}.}
\label{stopper coredrawing}
\end{figure}

The selected optimal configuration~\cite{Pilan} (Fig.~\ref{stopper coredrawing}) begins with four separate 40-mm-thick diluter slices (to allow free thermal expansion) made from Inconel~718, as described in Table~\ref{Incoprop}, which absorb part of the incoming beam energy. The remaining 400~mm is an absorber made from CuCr1Zr, as described in Table~\ref{Cuprop}, and this absorbs the remaining energy.

Age hardening was performed on the Inconel~718, while the CuCr1Zr underwent non-destructive ultrasonic testing to verify that there were no defects inside the material and to ensure that both materials would be able to absorb enough beam energy.

No active cooling is required for the beam stoppers as they are not intended to receive multiple pulses on a regular basis neither for a long periods. The stopper core is passively cooled via heat conduction to the stopper-core supports and marginally through radiation to the vacuum chamber~\cite{Pilan}.

\FloatBarrier
\subsection{Working principle: Demonstrator}
\label{Working principle - Demonstrator}
There are several ways to ensure personnel safety during operation of the beam stoppers \cite{SLAC,SPIRAL2,GARIS2,Fermilab}. CERN decided to differentiate machine protection and personnel protection by using two different sets of equipment, but these are combined into a single access-control system and associated with upstream bending magnet (BHZ).

Being classified as an ``Element Important for Safety,'' a beam stopper must lock the injection, transfer, or circulation of the beam to protect personnel from beam exposure using security functions according to CEI/IEC~61508, giving it a Safety Integrity Level of 3 (SIL3). This means that a risk-reduction factor greater than 1000 is required.\cite{CEI}

To meet these requirements, the beam stoppers work with the following conditions. In case of personnel access, beam stoppers must be in the in-beam position. In case of circulating beam or injection, the VETO signal is removed and the beam stoppers are moved out of the beam aperture. In case of intrusion, beam stoppers are automatically forced into the in-beam position. The beam stoppers work in fail-safe conditions, meaning that they will move into the in-beam position automatically in case of electrical, pneumatic, control, or vacuum failure. Furthermore, a redundancy rule is applied, meaning that except for a few beam lines, a minimum of two beam stoppers are installed in series.

One of the requirements of the new beam stoppers is a 30-year lifetime. To establish which of the existing beam stoppers had actually performed the most cycles, the number of pulses of the end-stroke switches of all the beam stoppers between 2008 to 2018 was extracted. It was found that each of the 17 beam stoppers in the accelerator complex performed between 177 and 3984 cycles, with an average of 1830. However, this is not necessarily the number of pulses that were received; it is only the number of cycles performed by each beam stopper, and this includes test cycles and cycles performed when access is given or under fail-safe conditions. Given the lifetime requirement and the history of cycles, the beam-stopper mechanism must last at least 30 years under 1000 cycles per year, the edge-welded bellows being the limiting component.

\begin{figure}[htb]
\centering
\includegraphics[scale=0.20]{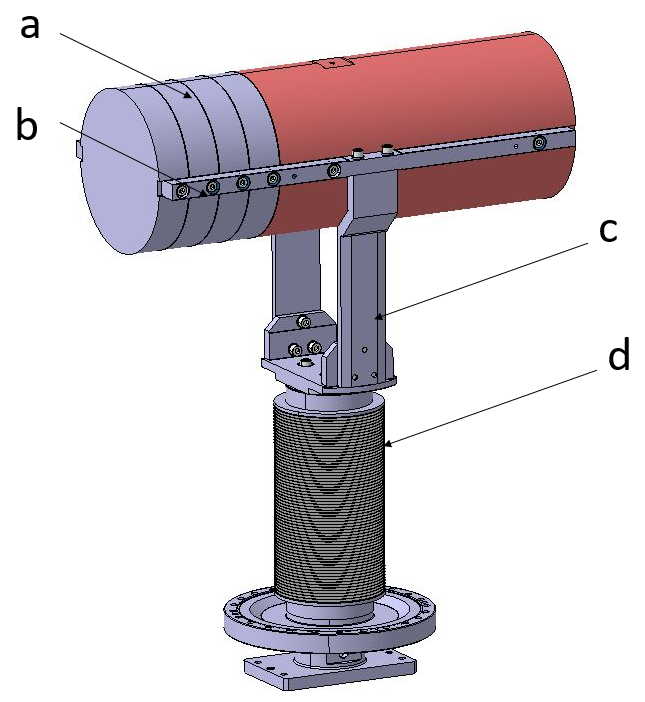}
\caption{3D model showing the stopper-core (a)materials supported by a fixing bar (b) on each side, assembled on a fork (c), and linked to an edge-welded bellows (d), allowing it to move under high vacuum.}
\label{stopper corefix1}
\end{figure}

Knowing the operational and weight requirements of the final stopper core, it was decided that its design should use a fork support (see Fig.~\ref{stopper corefix1}). In the out-beam position, the stopper core is located above the beam aperture, allowing the beam to pass through during normal operation, and it moves to the in-beam position during emergency scenarios or according to safety requirements. The fork is large enough to not obstruct the 156-mm beam aperture in normal operation.

\begin{figure}[ht]
\centering
\includegraphics[scale=0.38]{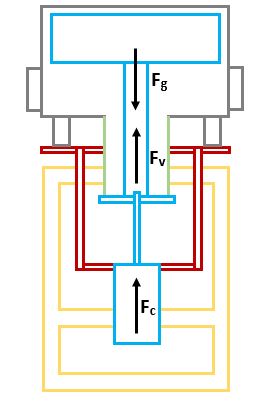}
\caption{Schematic of the forces exerted on the beam-stopper assembly in case of pneumatic failure. Here, $F_{\mathrm{g}}$ represents the weight of the stopper core, $F_{\mathrm{v}}$ represents the vacuum force exerted on the outer edge-welded bellows, and $F_{\mathrm{c}}$ represents friction forces.}
\label{BSscheme}
\end{figure}

To fulfill the safety rules, the weight of the stopper-core assembly has to overcome the vacuum force and the friction and bellows forces in any emergency or normal scenario (Fig.~\ref{BSscheme}). This is helped---if operating---by a Festo\textregistered DSBC-100-180 linear pneumatic cylinder.

\begin{figure}[htb]
\centering
\includegraphics[scale=0.70]{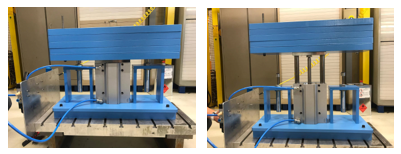}
\caption{Photographs of the demonstrator, showing the in-beam (left) and out-beam (right) positions.}
\label{Demonstrator}
\end{figure}

A demonstrator was built (Fig.~\ref{Demonstrator}) to verify whether the beam stopper fulfilled all of the functional requirements, to check the resulting forces, and to obtain a quantitative value for the friction forces. The demonstrator was also used to verify the stroke time, test the pneumatic parameters, and to obtain qualitative information about the end-stroke shock absorbers.

\subsection{Standard adaptable design}
\label{Standard adaptable design}
\begin{figure}[htb]
\centering
\includegraphics[scale=0.68]{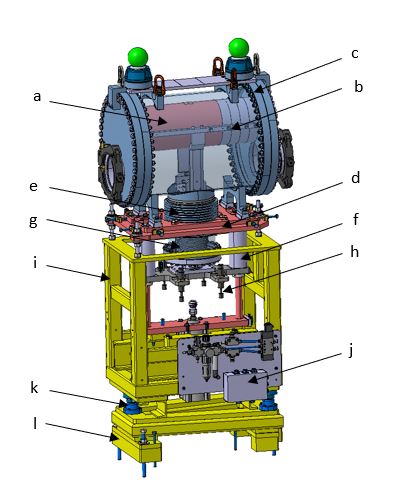}
\caption{3D model with arrows indicating the main components of the beam stopper as designed, the core in "out" position, beam coming from the right side of the picture.}
\label{BSarrow}
\end{figure}

The new beam-stopper assembly consists of the following sub-systems: the stopper core (labelled ``a'' in Fig.~\ref{BSarrow}), which is held by the fork (b) inside a vacuum vessel (c); the fork supports a hydroformed bellows (e), which is itself connected to the support plate for the shock absorbers (h), with the guiding systems allowing the stopper core to move (f). All these sub-assemblies are fixed to an adjustment table (d). The support is divided into a lower (l) and an upper plug-in (i), which are connected using an isostatic ball-mounting system (k). The pressurized pneumatic cylinder is fixed onto the guiding system and is powered by a patch panel (j) attached to the upper plug-in. The stopper-core assembly is damped through shock absorbers on the upper plug-in.

\begin{figure}[htb]
\centering
\includegraphics[scale=0.35]{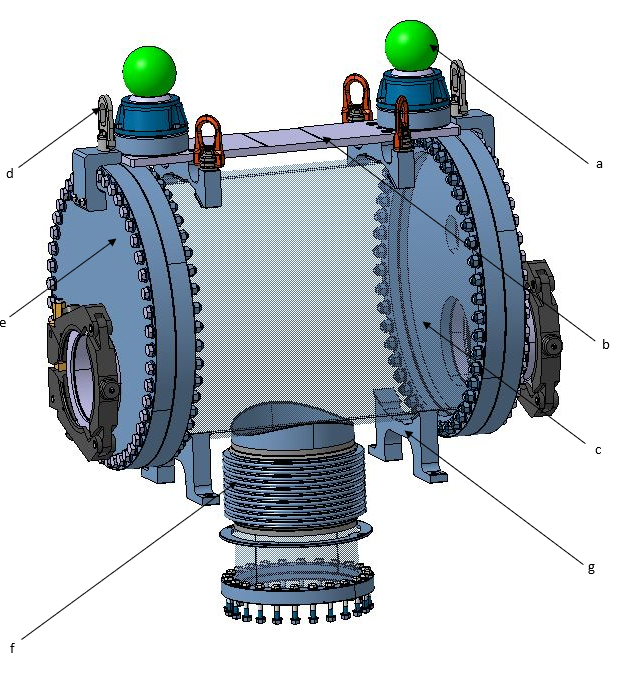}
\caption{3D model with arrows showing the main parts of the vacuum-vessel assembly.}
\label{tankassy}
\end{figure}

CERN guidelines for equipment operating under vacuum require an operational pressure between $5 \times 10^{-3}$ and $5 \times 10^{-8}$~mbar, an average pumping speed of 100~l\,s$^{-1}$, and an outgassing rate limited to between $5 \times 10^{-5}$ and $5 \times 10^{-6}$~mbar\,l\,s$^{-1}$, depending on where the beam stopper is installed. This means that the vacuum vessel must ensure high vacuum tightness. The vacuum vessel is a large chamber (labelled ``c'' in Fig.~\ref{tankassy}) made of AISI 304L stainless steel. On each side, ultra-high vacuum ConFlat-type flanges are secured using AISI 316LN 3D-forged screws on the tubes (e). The hydroformed bellows (f) is directly welded onto the vacuum vessel, allowing stretching of the cylinder. To tilt the vacuum vessel, shims can be installed underneath it (g). The top part (b) of the vacuum vessel is used to install geodesic (a) and lifting (d) instruments. A viewport is installed on the upstream flange of each beam stopper to allow visual inspection of the stopper core.

\begin{figure*}[htb]
\centering
 \includegraphics[width=0.9\textwidth]{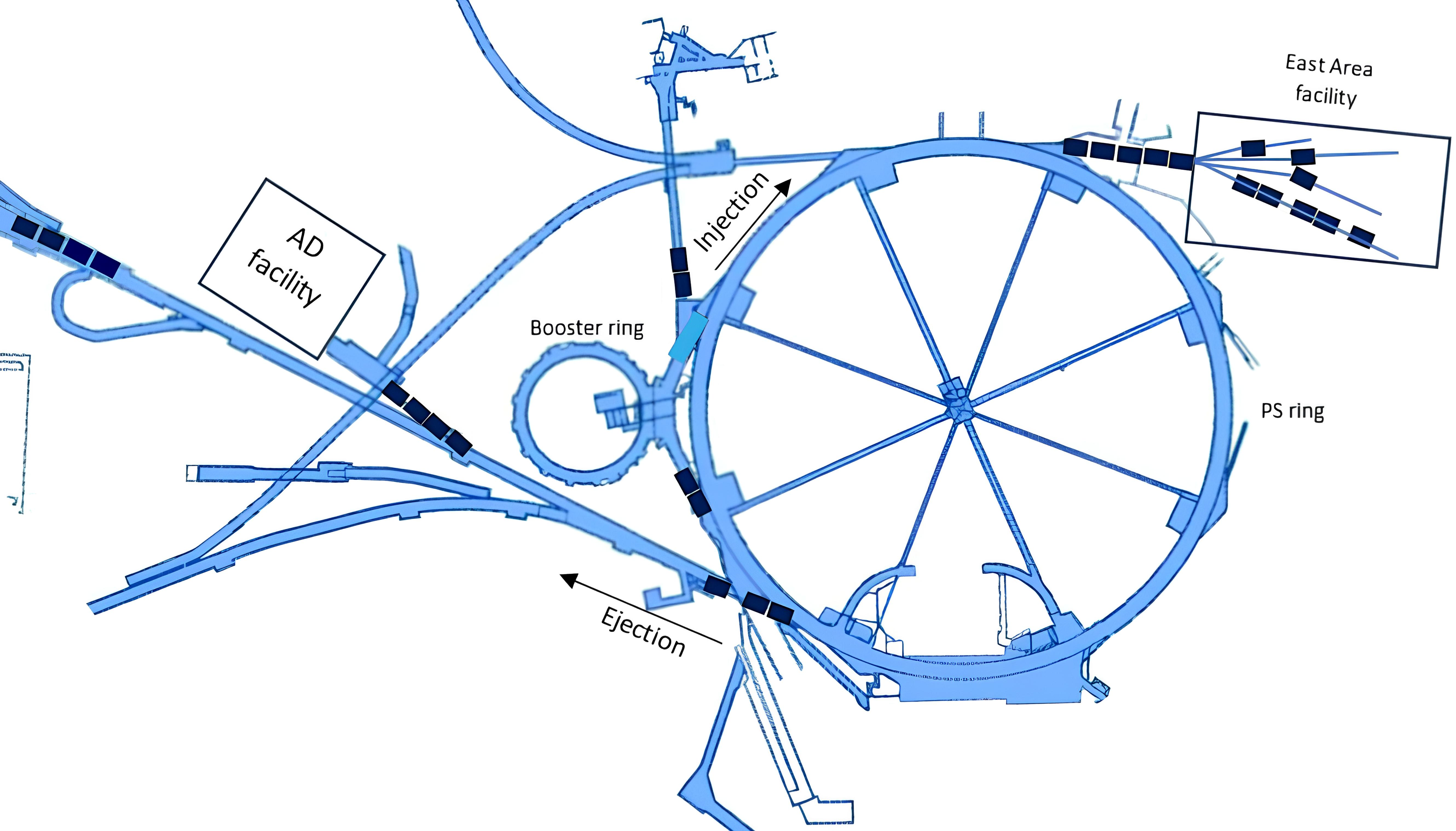}
\caption{Diagram showing the positions of the 28 beam stoppers (dark blue boxes) to be installed during Long Shutdown~3.}
\label{LaypostLS2}
\end{figure*}

The installation of the 28 new beam stoppers is expected to be carried out over several years in different facilities and injector lines. A total of 18 beam stoppers were installed before the start of the operational run in 2021. Ten more beam stoppers will be installed during Long Shutdown~3 (2027-2029). To increase ease of maintenance, it is important to have a standard beam-stopper base, which must be adapted to and compatible with the 28 installation points in the complex indicated in Fig.~\ref{LaypostLS2}.

\begin{figure}[htb]
\centering
\includegraphics[width=1.7cm, height=3.9cm]{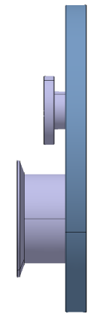}
\includegraphics[width=1.7cm, height=3.9cm]{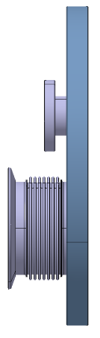}
\includegraphics[width=1.7cm, height=3.9cm]{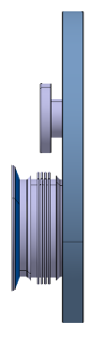}
\caption{3D models showing side views of three different flange types, the aperture and design of which vary according to the specific beam line.}
\label{Flanges}
\end{figure}

\begin{figure}[htb]
\centering
\includegraphics[scale=0.22]{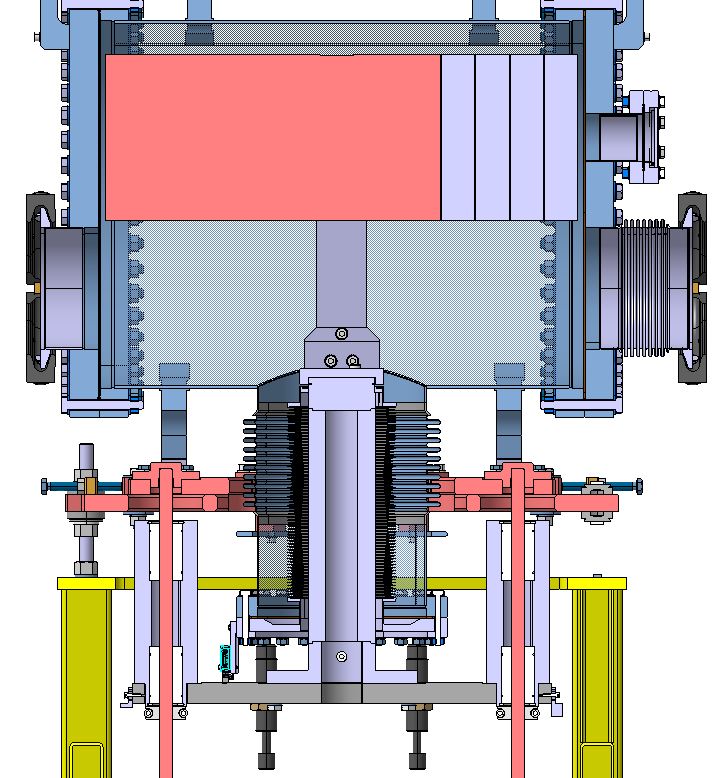}
\caption{Cross-sectional view of a 3D model showing the four linear bearings (in white) installed on the guiding rod (in red) each side of the bellows.}
\label{stopper corefix2}
\end{figure}

The vacuum vessel has a compact and standardized flange-to-flange length. The extremity flanges are adaptable according to their respective beam lines. A 150-mm flange diameter is required for the PS complex, while a 159-mm flange diameter is required for other facilities (Fig.~\ref{Flanges}). This involves several constraints. In the out-beam position, the stopper core must leave a free aperture for the beam that is equal in size to that seen in the existing beam stoppers. For a stopper-core diameter of 200~mm, the distance between the stopper-core center and the beam line must be greater than 175~mm, as it was before 2019. To minimize the scattering of secondary particles, the beam axis and the stopper-core center must be aligned within $\pm$2~mm in the in-beam position. This is achieved by the precise fabrication tolerances of the stopper-core assembly in the vacuum vessel, which is linked together by a guiding system equipped with linear bearings with a maximum eccentricity of 15~$\upmu$m (Fig.~\ref{stopper corefix2}).

\begin{figure}[htb]
\centering
\includegraphics[scale=0.26]{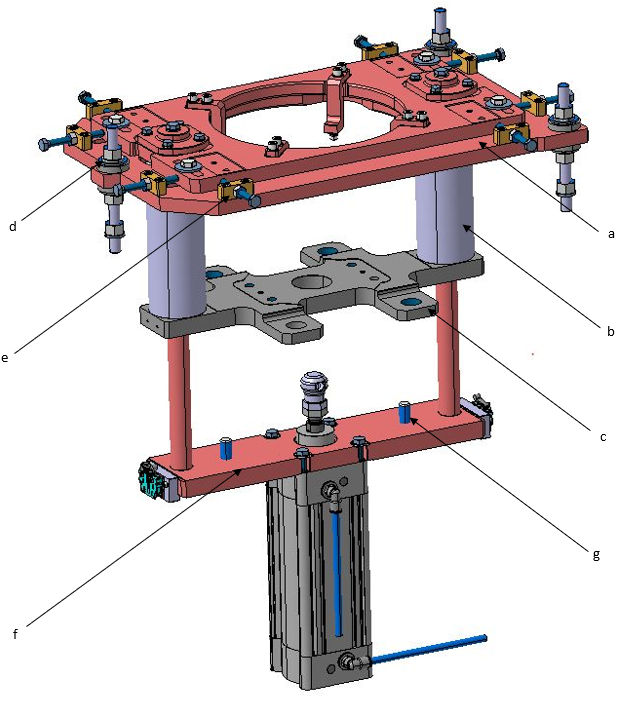}
\caption{3D model of guiding system and alignment table.}
\label{Guidingsyst}
\end{figure}

\begin{figure}[htb]
\centering
\includegraphics[scale=0.18]{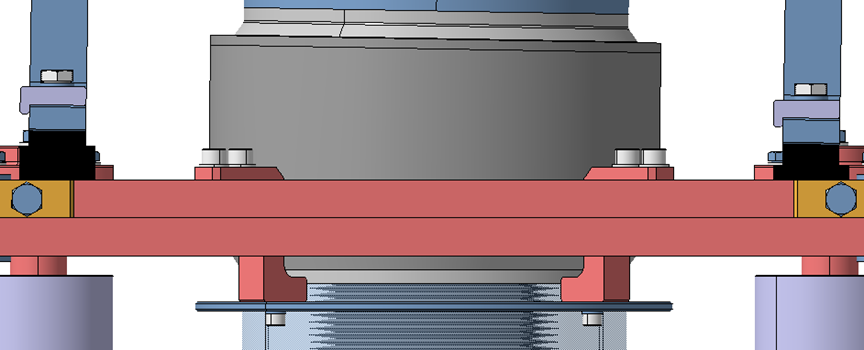}
\includegraphics[scale=0.18]{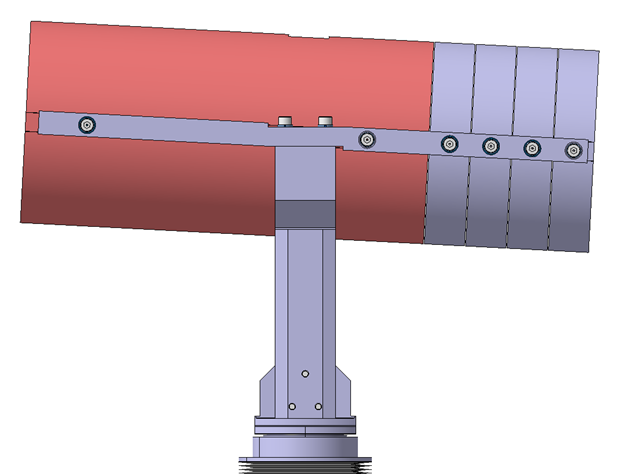}
\caption{3D view showing the vacuum-vessel shims (left) and the inclined machined surface of the fixing support (right).}
\label{shims}
\end{figure}

Each beam stopper is equipped with a standard alignment table (labelled ``a'' in Fig.~\ref{Guidingsyst}) to maintain coaxiality within $\pm$0.5~mm between the flanges and the beam line. The vacuum vessel and stopper core can be moved using the push and pull parts (e) for two-axis horizontal adjustment and one rotation on the horizontal plane. Three threaded screws and concave/convex washers (d) allow the vacuum-vessel and stopper-core height to be adjusted by $\pm$30~mm and to adjust the tilt and slope angle of the vacuum vessel after the pre-setting illustrated in Fig.~\ref{shims}.

The guiding system is also used to maintain the guiding shafts (``b'' in Fig.~\ref{Guidingsyst}) and the shock-absorber plate (c). The pneumatic actuator is installed on the base (f). The shock absorbers have been dimensioned to smoothly absorb the forces on the stopper core during normal and emergency scenarios. Each guiding shaft has a ring that serves as an end stop for the out-beam position. Each also has two screws that work as stops for pneumatic-cylinder replacement (g).

The beam-stopper assemblies must cope with tunnel gradients from $-3^\circ$ to $+3^\circ$. The technical solution presented here allows this to be set in increments of $1^\circ$. To this end, shims (shown in black in Fig.~\ref{shims}) are positioned below the lower support of the beam-stopper vacuum vessel to tilt it, and the hydroformed bellows allows this tilt. The fixing support of the stopper core (as shown in the right-hand panel of the figure) can have an inclined machined surface. By setting the same slope angle on the vacuum vessel and the stopper core, they remain parallel to adapt to the slope of the beam line. The purpose of this solution is to keep the guiding system vertical so as not to induce additional forces on the mechanism.

Each beam stopper is equipped with a standard upper plug-in and an adaptable lower plug-in to position it isostatically and allow it to be removed quickly, to comply with the ``as low as reasonably achievable'' (ALARA) principle of radioprotection. The lower plug-in can be produced to different designs to accommodate beam heights between 1 and 2~m.

\begin{figure}[htb]
\centering
\includegraphics[scale=0.46]{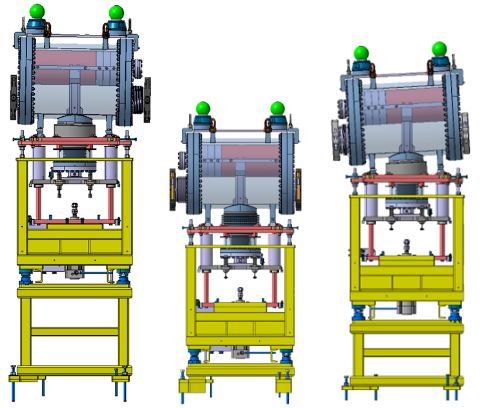}
\caption{Three different example 3D models illustrating different possibilities of lower plug-in heights, flanges, and shims.}
\label{Beamconfig}
\end{figure}

Through the described combination of adaptable flanges, shims, different lower plug-ins, table settings, and redundancy, beam stoppers of this design can be installed in every position of the complex in which they are required (Fig.~\ref{Beamconfig}).

\begin{figure}[htb]
\centering
\includegraphics[width=0.4\textwidth]{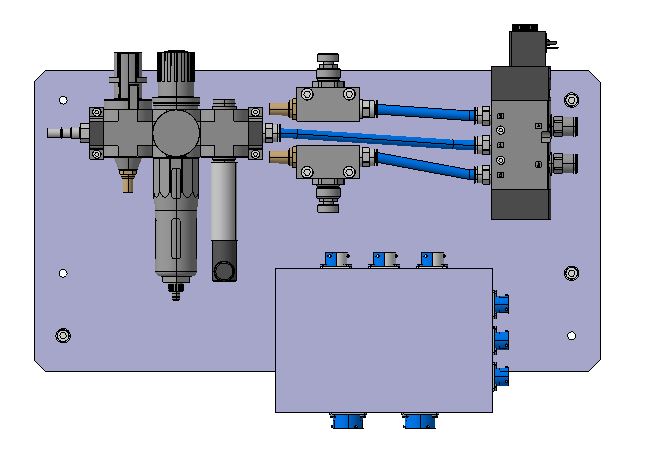}
\caption{3D model of the patch panel.}
\label{Patch}
\end{figure}

\subsection{Pneumatic and electrical systems}
\label{Pneumatic and electrical systems}
\begin{figure}[htb]
\begingroup\centering
\includegraphics[scale=0.5]{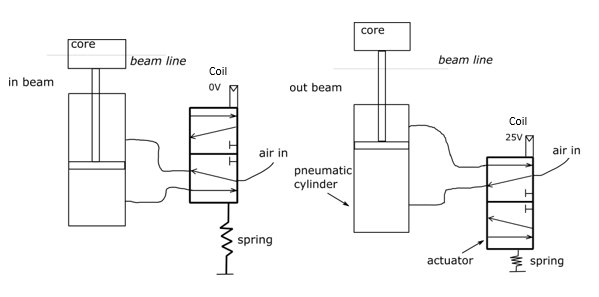}
\caption{Schematic pneumatic circuit of the beam stopper in the in-beam (left) and out-beam (right) positions, showing the solenoid valve and pneumatic actuator.}
\label{patchscheme}
\endgroup
\end{figure}

The patch panel (Fig.~\ref{Patch}) supports the pneumatic circuit for actuating the pneumatic cylinder, as well as an electrical box. Fig.~\ref{patchscheme} shows a schematic of the pneumatic circuit used in the beam stopper. When the actuator is pressurized, the stopper core stays is in the out-beam position. When the 5/2 monostable solenoid valve is actuated, the mechanical spring brings the stopper core back to the in-beam position. To maintain fail-safe operation, in case of various failure conditions, the stopper core is moved to the in-beam position for safety. If there is a power cut, the stopper core will move to the in-beam position due to the mechanical spring of the electro-valve pulling the solenoid valve; this will dump the pressure inside the actuator. If there is an air-pressure drop in the circuit, the weight of the stopper core will cause it to move to the in-beam position, even if the pneumatic actuator is not pressurized.

\begin{figure}[htb]
\centering
\includegraphics[width=0.25\textwidth]{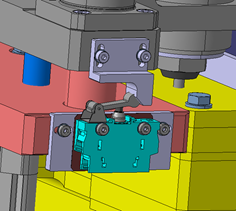}
\caption{In-beam position switch.}
\label{Switch}
\end{figure}

The beam stoppers are equipped with two in-beam position switches (Fig.~\ref{Switch}), fixed on either side of the bottom plate of the guiding system: one for the control system and one for the access safety system. There is also an out-beam switch fixed onto the shock plate.

Taking into account the use of these control components in a radiation environment, they are positioned far from the beam stopper, and the electronic and pneumatic components installed on the beam stopper---along with the polymer-based seals---are radiation resistant to the values expected in the different locations (up to the MGy level).

\subsection{Mechanical assessments of the system}
\label{Mechanical assessments of the system}
\begin{figure}[htb]
\centering
\includegraphics[scale=0.50]{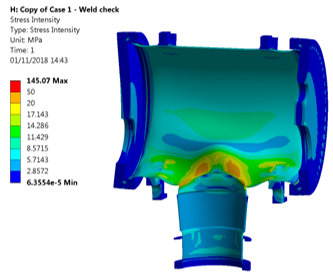}
\includegraphics[scale=0.65]{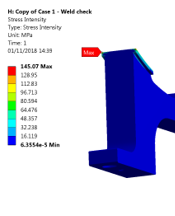}
\caption{Vacuum vessel pressure static structural simulation mapping illustrating the position of the peak stress intensity.}
\label{tankpressuresim}
\end{figure}

Ansys Structural simulations were performed to validate the mechanical behavior of the vacuum vessel under pressure. The results (Fig.~\ref{tankpressuresim}) show a peak stress intensity of 145~MPa, which is within the standard required for pressurized vessels according to EN~13445-3.

The welds were divided into sections and analyzed according to the NF EN 1993-1-8 standard, since larger stresses are observed at these points. The requirements for strength, serviceability, and durability of structures are defined by:
\begin{equation}
[\sigma_{\perp}^{2}+3(\tau_{\perp}^{2}+\tau_{\parallel})]^{0.5}\leq\frac{f_{\mathrm{u}}}{\beta_{\mathrm{w}}\gamma_{\mathrm{M2}}},
\end{equation}
and
\begin{equation}
    \sigma_{\perp}\leq\frac{0.9f_{\mathrm{u}}}{\gamma_{\mathrm{M2}}},
\end{equation}
where $\sigma_{\perp}$ is the nominal stress perpendicular to the groove, $\tau_{\perp}$ is the shear stress perpendicular to the axis of the weld, $\tau_{\parallel}$ is the shear stress parallel to the axis of the weld, $f_{\mathrm{u}}$ is the nominal ultimate tensile strength of the welded component, $\gamma_{\mathrm{M2}}$ is the partial safety factor for the resistance of welds, and $\beta_{\mathrm{w}}$ is the appropriate correlation factor.

The results show von-Mises-equivalent stresses of 30, 50, and 65~MPa for the welds of the flanges, the upper support, and the lifting supports, respectively, which are lower than the 100-MPa limit defined by the NF EN 1993-1-8 standard.

\begin{figure}[htb]
\centering
\includegraphics[scale=0.5]{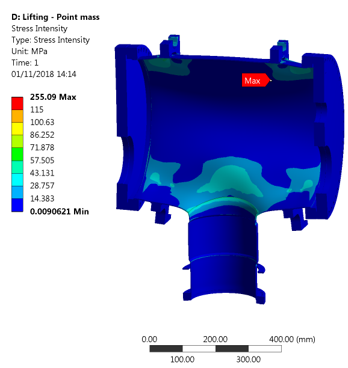}
\caption{Results of a structural simulation of the vacuum-vessel static lifting case, illustrating the position of the peak stress intensity.}
\label{tankliftsims}
\end{figure}

Another simulation was performed for the lifting of the beam stopper for all handling processes such as transport and installation (Fig.~\ref{tankliftsims}). A conservative approach was used, assuming a 1.5$g$ acceleration and a total mass of 600~kg (instead of 535~kg). The lifting was performed using four points on the top of the vacuum vessel. The peak stress was found to reach 255~MPa in the welded area of the lifting support. Even though the stress limit according to EN~1993-1-8 is 110~MPa, the welds have been validated by the simulations above. Additional elasto-plasticity calculations with 3$g$ acceleration were performed, showing acceptable results.

\begin{figure}[htb]
\centering
\includegraphics[scale=0.8]{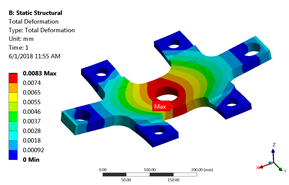}
\includegraphics[scale=0.8]{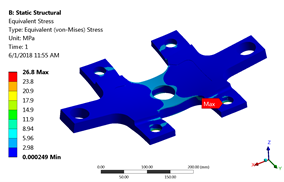}
\caption{Simulated stresses (left) and deformations (right) in the shock-absorber plate.}
\label{Shockplatesims}
\end{figure}

The stresses and deformations caused by the shock-absorber plate on the guiding system were calculated using a static structural simulation, as illustrated in Fig.~\ref{Shockplatesims}. A worst-case scenario with a $3^\circ$ tilt between both systems was considered, adding a lateral force and a moment to the bottom plate. The maximum total deformation of the bottom plate was found to reach 0.0083~mm, while the maximum equivalent von Mises stress was 26.8~MPa, lower than the yield strength at $20^\circ$C (180~MPa).

\section{Installation of new beam stoppers}
\label{Installation of new beam stoppers}

\subsection{Main assembly steps}
\label{Main assembly steps}
The total weight of the stopper core, which is composed of Inconel~718 and CuCr1Zr, is 160~kg. This means that a special spreader is needed for the insertion of the stopper core into the vacuum vessel during the assembly phase. This spreader is made from S235JRG2 steel with a hook welded to it to enable lifting of the stopper core from above its center of gravity.

Structural simulations were performed to simulate the equivalent stresses and deformations of the spreader when lifting the stopper core. The simulations used a safety factor of 3 by simulating 3 times the actual weight of the stopper core. The results show a maximum equivalent stress of 120~MPa for a deformation of 2.1~mm.

\begin{figure}[htb]
\centering
\includegraphics[scale=0.80]{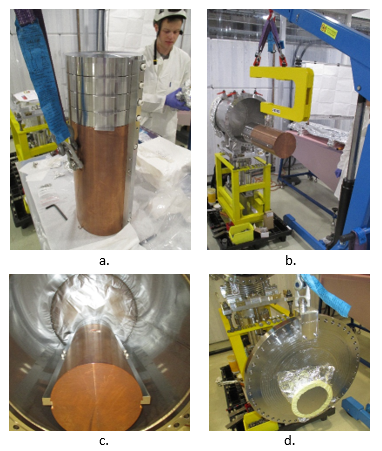}
\caption{Stopper-core installation process: (a)~attachment; (b)~lifting with the spreader and installation into the vacuum vessel; (c)~attachment to the fork; (d)~closing the vacuum vessel.}
\label{coreinstall}
\end{figure}

The stopper core is assembled, lifted, and inserted into the vacuum vessel in a clean room and then fixed to the previously installed fork. The vacuum vessel is then closed with the flanges, as shown in Fig.~\ref{coreinstall}.

\subsection{Removal of old beam stoppers}
\label{Removal of old beam stoppers}
\begin{figure}[htb]
\centering
\includegraphics[scale=0.024, angle=270]{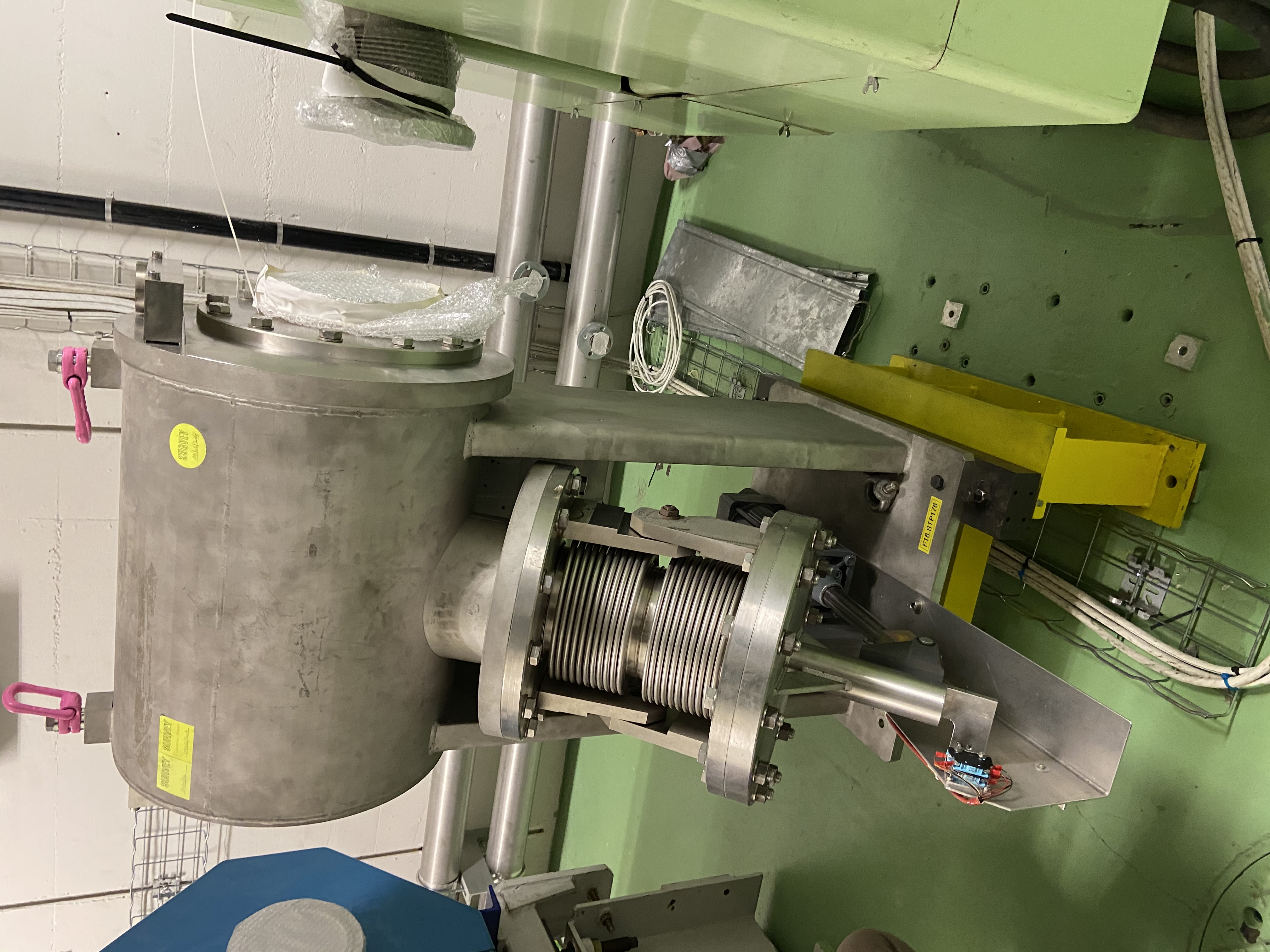}
\includegraphics[scale=0.024, angle=270]{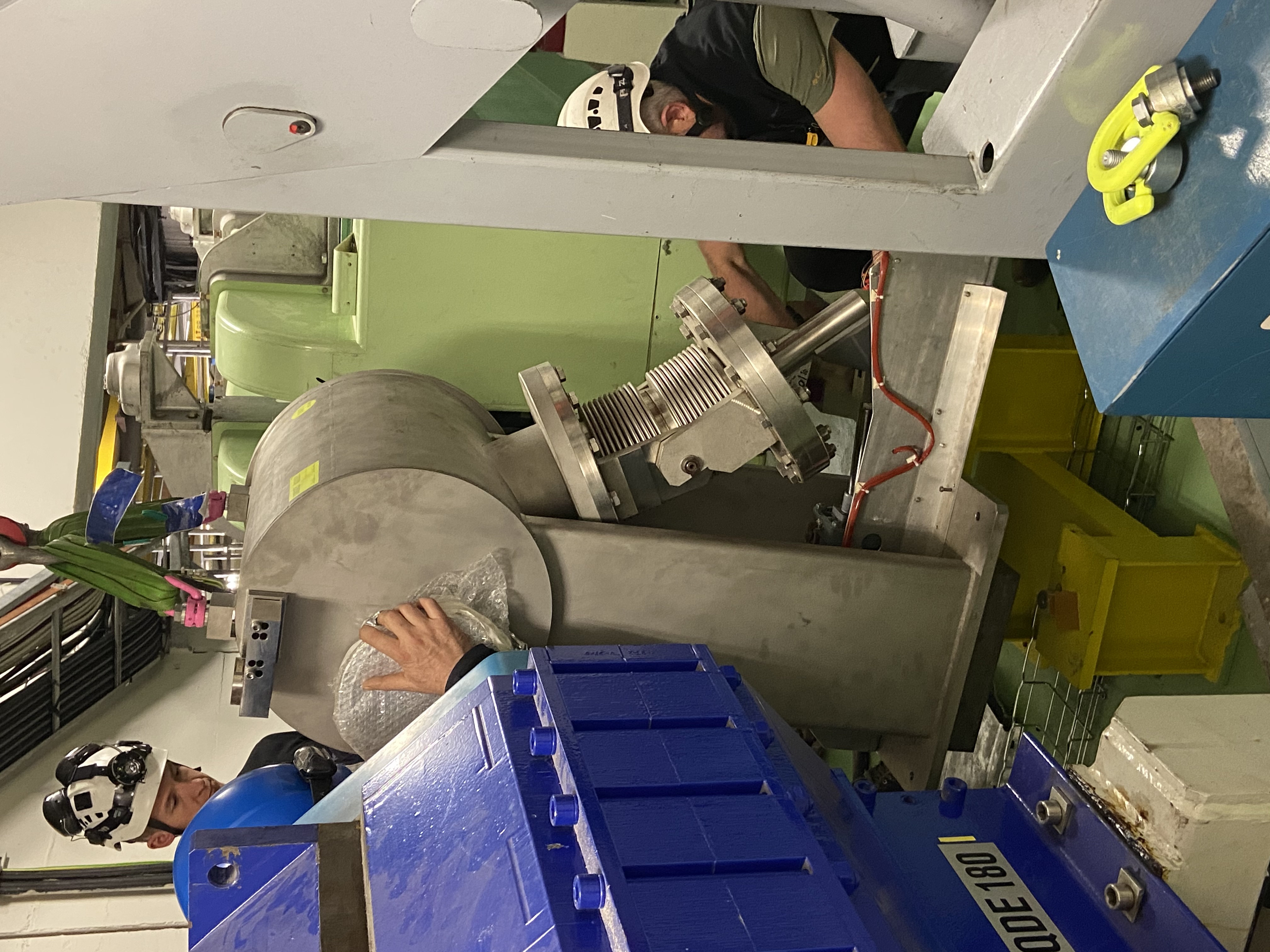}
\includegraphics[scale=0.024, angle=270]{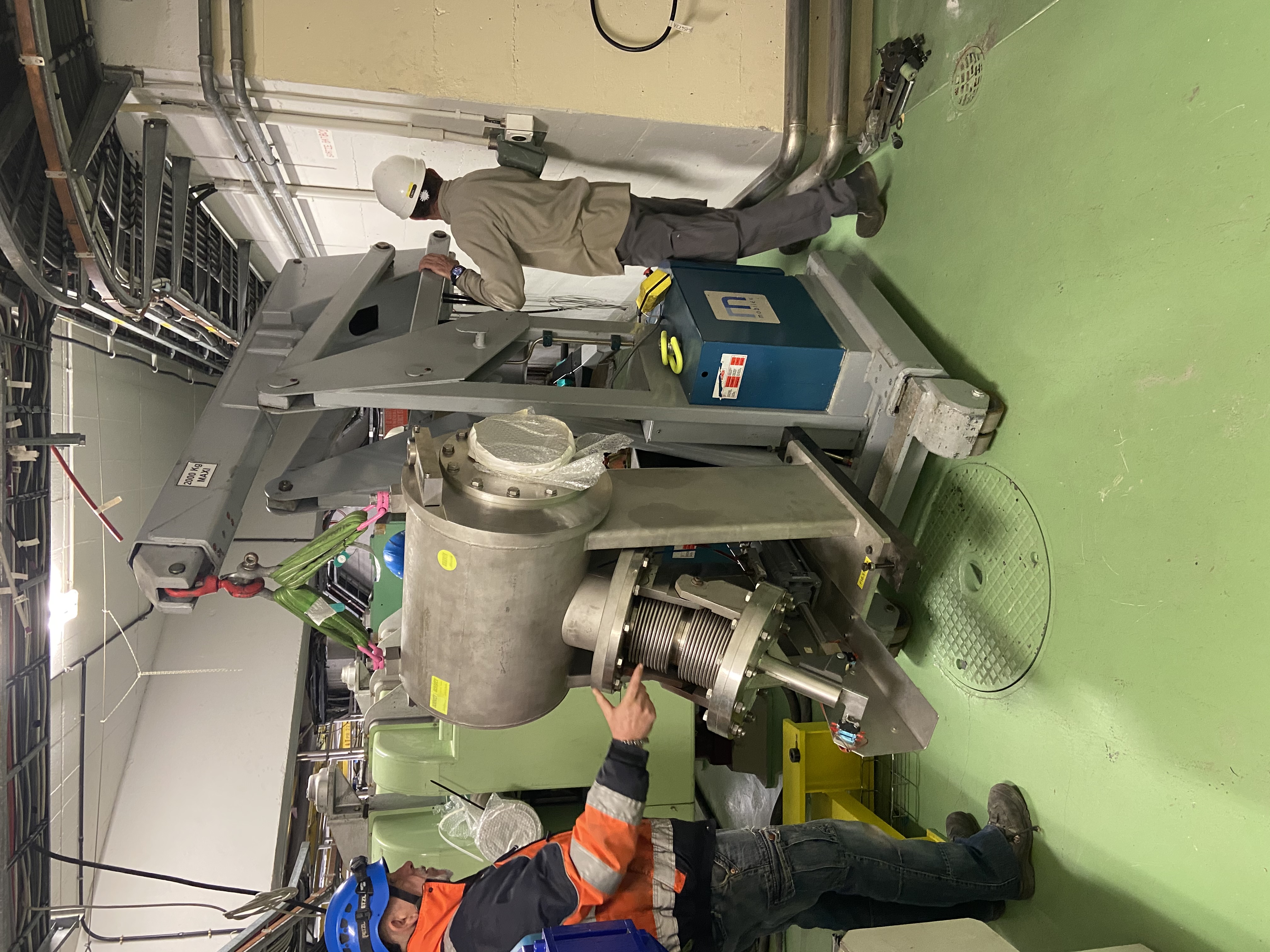}
\caption{Removal of Fig.~\ref{Bemact}(b)-type beam stopper from CERN transfer line.}
\label{removal}
\end{figure}

The first scheduled removal of old beam stoppers was performed during LS2. We take the example of the removal of a Fig.~\ref{Bemact}(b)-type beam stopper. Photographs of this process are shown in Fig.~\ref{removal}. The first step in this process is disconnection of the electrical and/or pneumatic systems according to their different designs; this is followed by lifting and extraction from the tunnels toward the surface. The extracted beam stoppers were transported to the radioactive-waste storage facility at CERN. The main challenge was the physical difficulty in extracting beam stoppers that resulted from surrounding equipment or the low heights of transfer lines in several tunnels.

\subsection{Installation and maintenance of the new beam stoppers}
As noted, the design is standard but adaptable, and the plug-ins may be different in terms of dimensions (depending on the installation point) but similar in terms of geometry. The standard shape of the vacuum vessel makes the lifting process identical in each case, with the same four lifting points. Finally, the isostatic ball-mounting system allows quick installation of the beam stoppers on their lower plug-ins. This plug-in system allows very rapid removal or exchange of a beam stopper if there is a need for corrective maintenance.

\begin{figure}[htb]
\centering
\includegraphics[scale=0.03, angle=270]{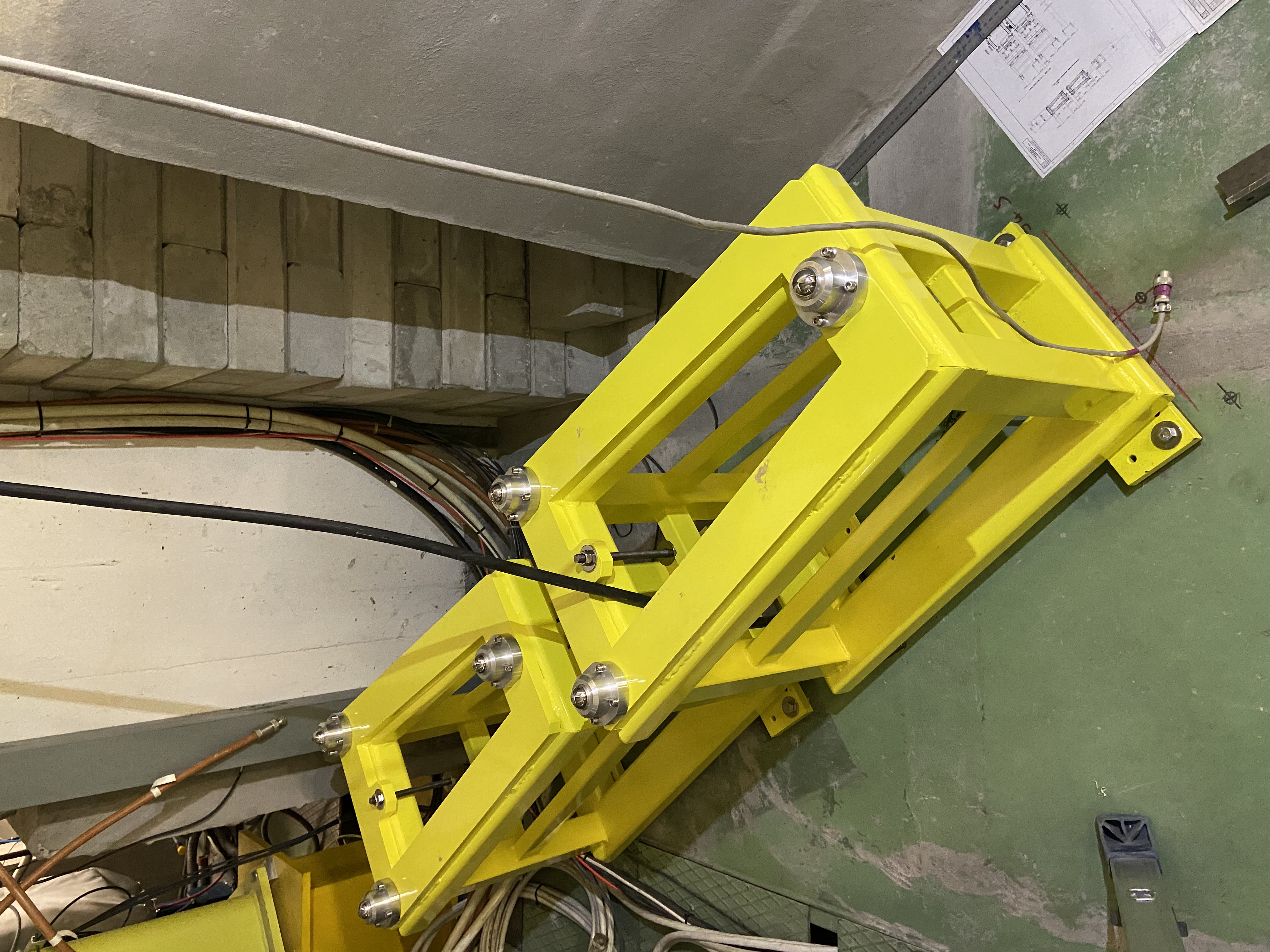}
\includegraphics[scale=0.03, angle=270]{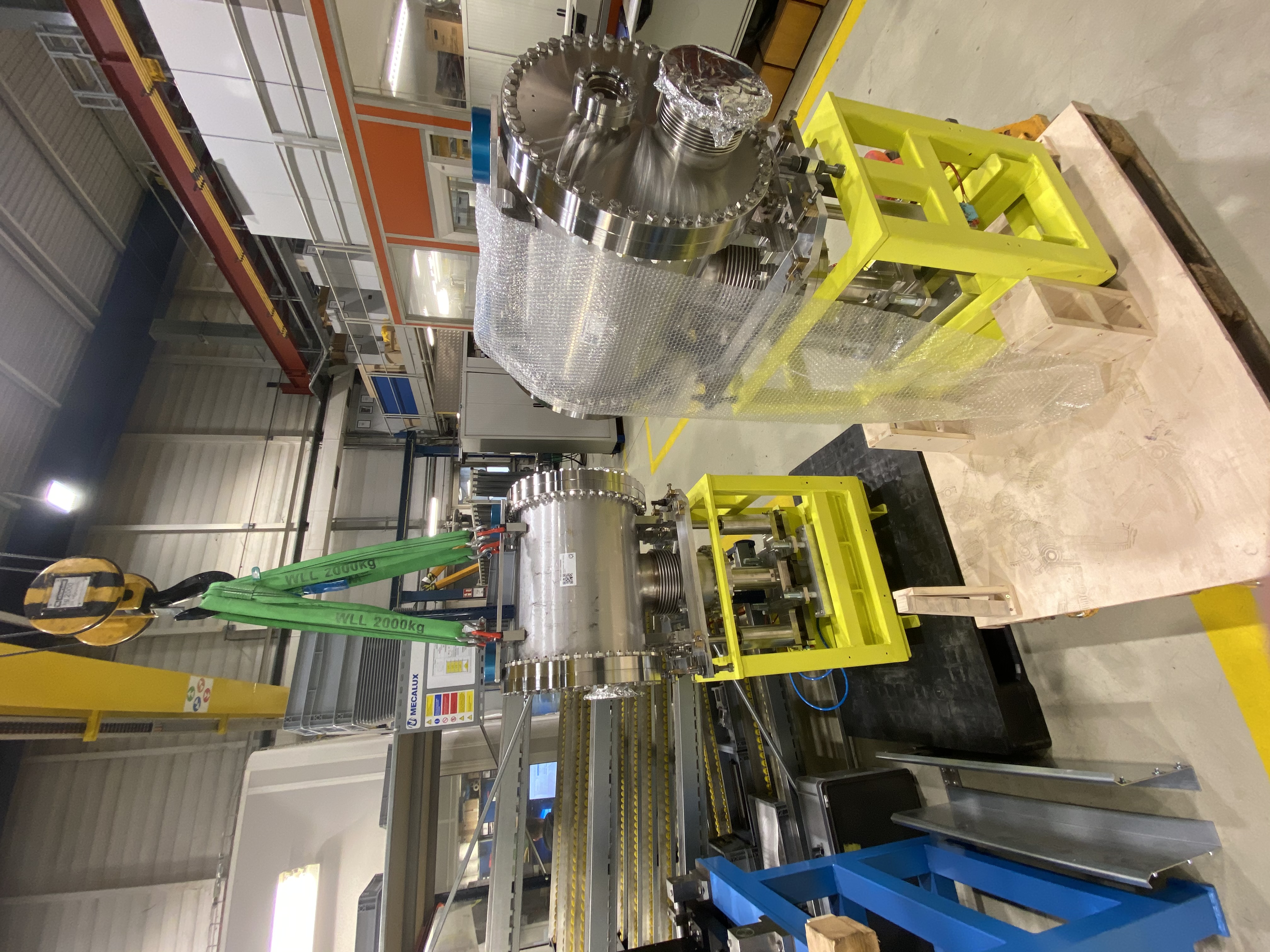}
\includegraphics[scale=0.03, angle=270]{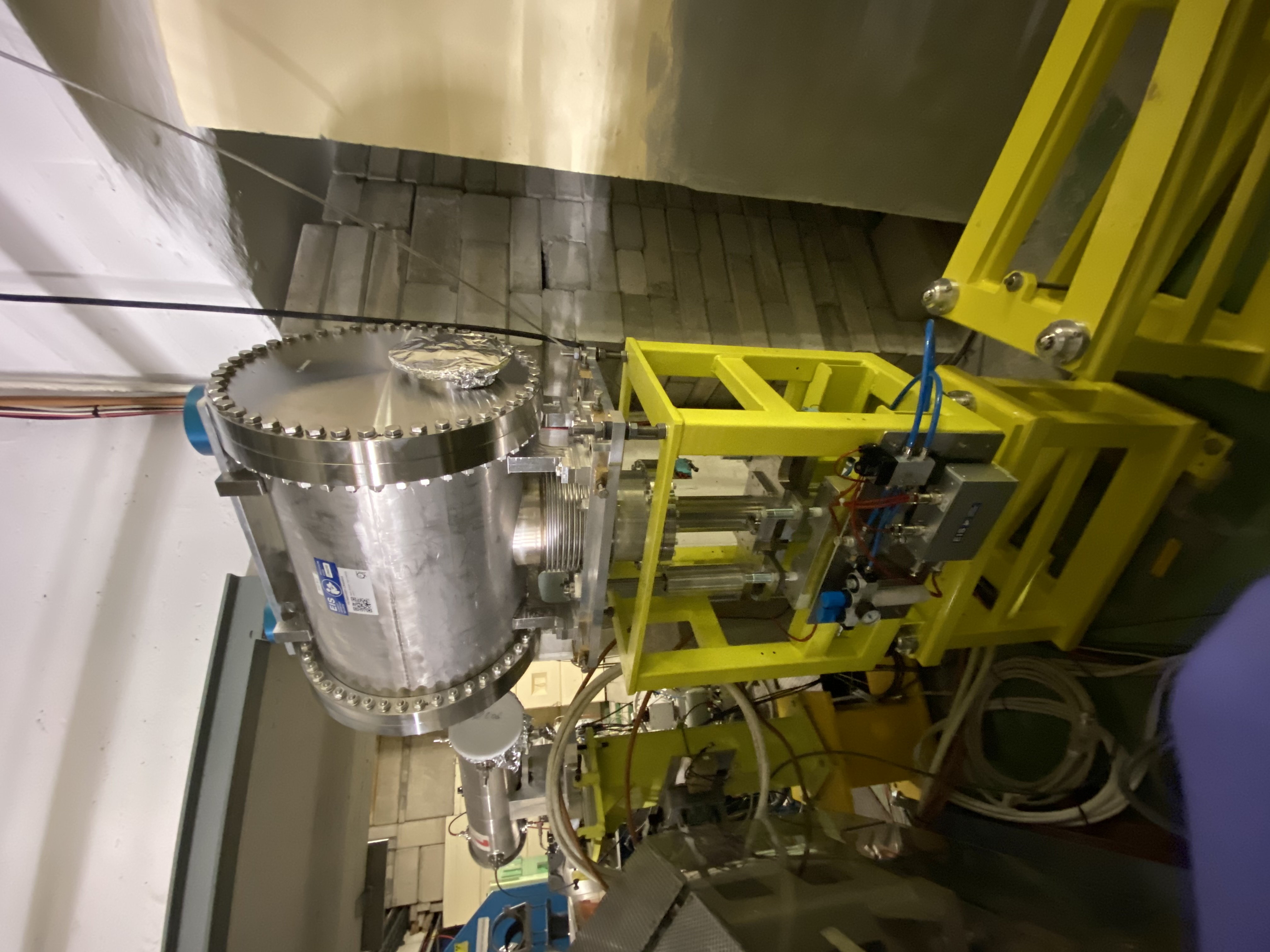}
\includegraphics[scale=0.03, angle=270]{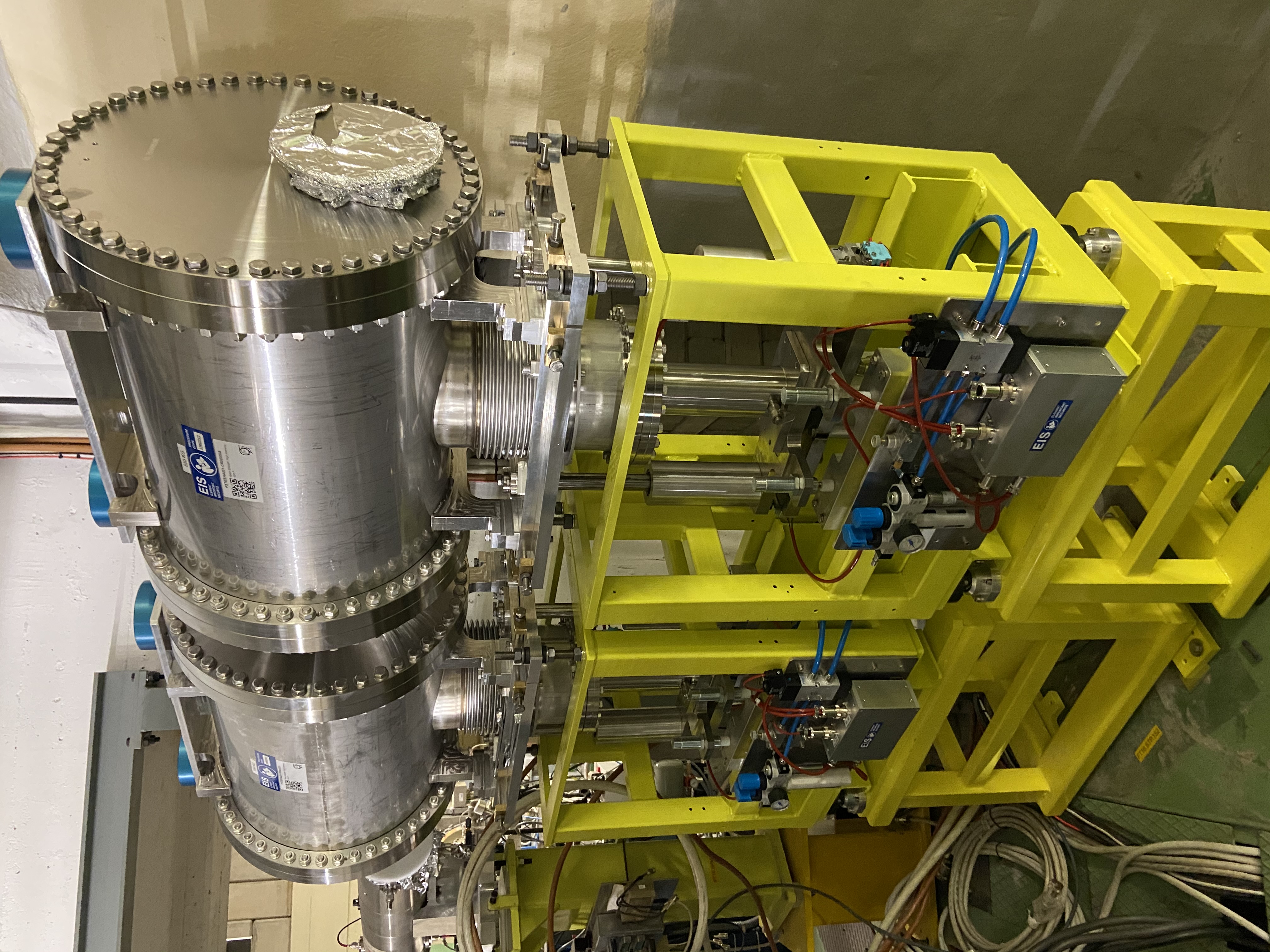}
\caption{Steps in the installation of the new beam stoppers: steps 1--4 are shown in the top left, top right, bottom left, and bottom right panels, respectively.}
\label{Install}
\end{figure}

Only four steps are required to install a beam stopper, as shown in Fig.~\ref{Install}; these are as follows. (1)~Fixing the lower plug-in to the ground. (2)~Lifting the beam stopper. (3)~Positioning the beam stopper on the lower plug-in. (4)~Making electrical and pneumatic connections. After this, CERN perform the alignment and connection between the upstream and downstream equipment.

\begin{figure}[htb]
\centering
\includegraphics[width=0.48\textwidth]{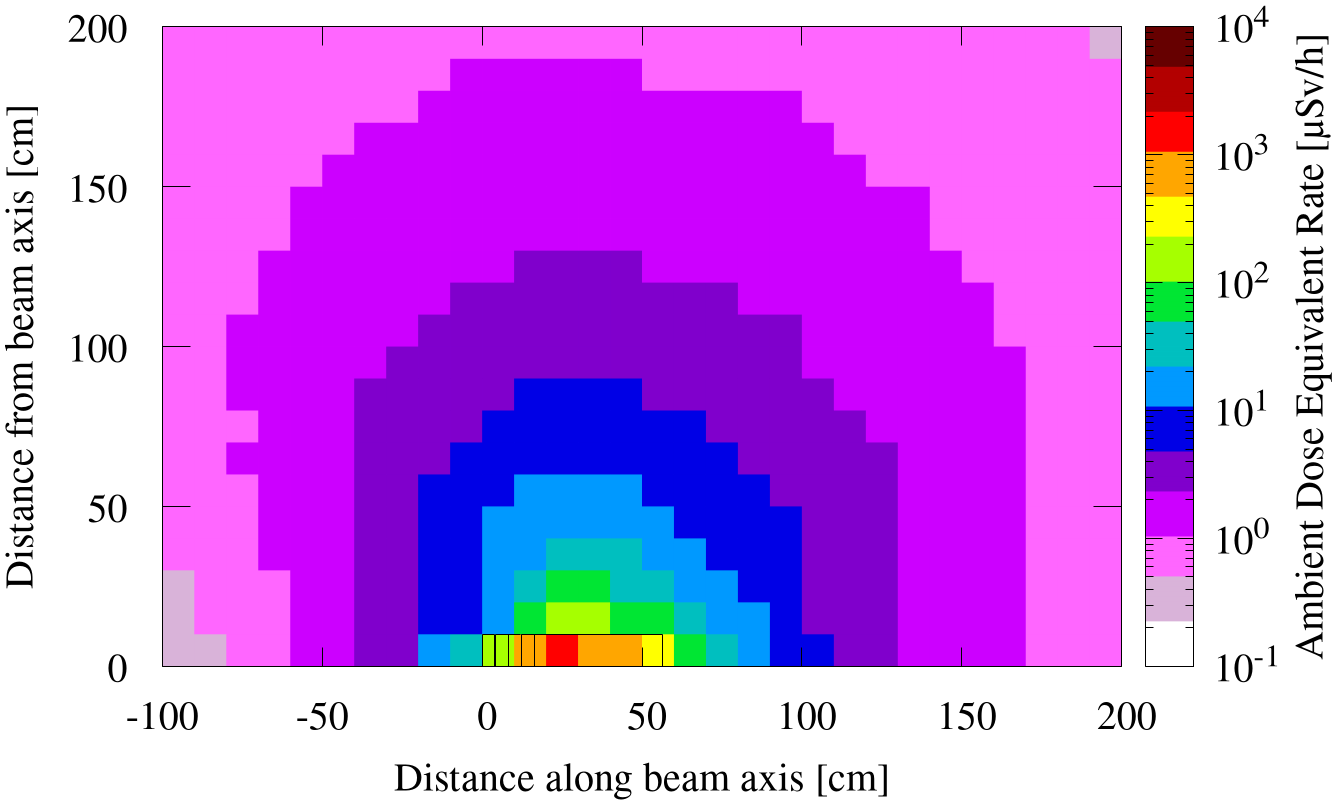}
    \caption{Residual ambient dose equivalent rate around the beam stopper as evaluated with FLUKA4-3.4. For the Monte Carlo simulations, an irradiation time of 15 pulses (as specified in Table~\ref{Beamparam}) and a cool-down time of 2~months were considered.}
    \label{residual-dose-rate}
\end{figure}

The residual ambient dose equivalent rates must be estimated for the work-dose planning of maintenance work on the beam stoppers. Fig.~\ref{residual-dose-rate} shows the results of these calculations, as estimated using FLUKA4-3.4 \cite{FLUKA1, FLUKA2, FLUKA3}, around the beam stopper after it has received 15 pulses as specified in Table~\ref{Beamparam} followed by a cool-down period of 2~months. This cool-down time corresponds to maintenance activities during a regular technical stop at the end of an operational year. It was found that the residual ambient dose equivalent rates allows for maintenance work.

\begin{figure}[htb]
\centering
\includegraphics[scale=0.50]{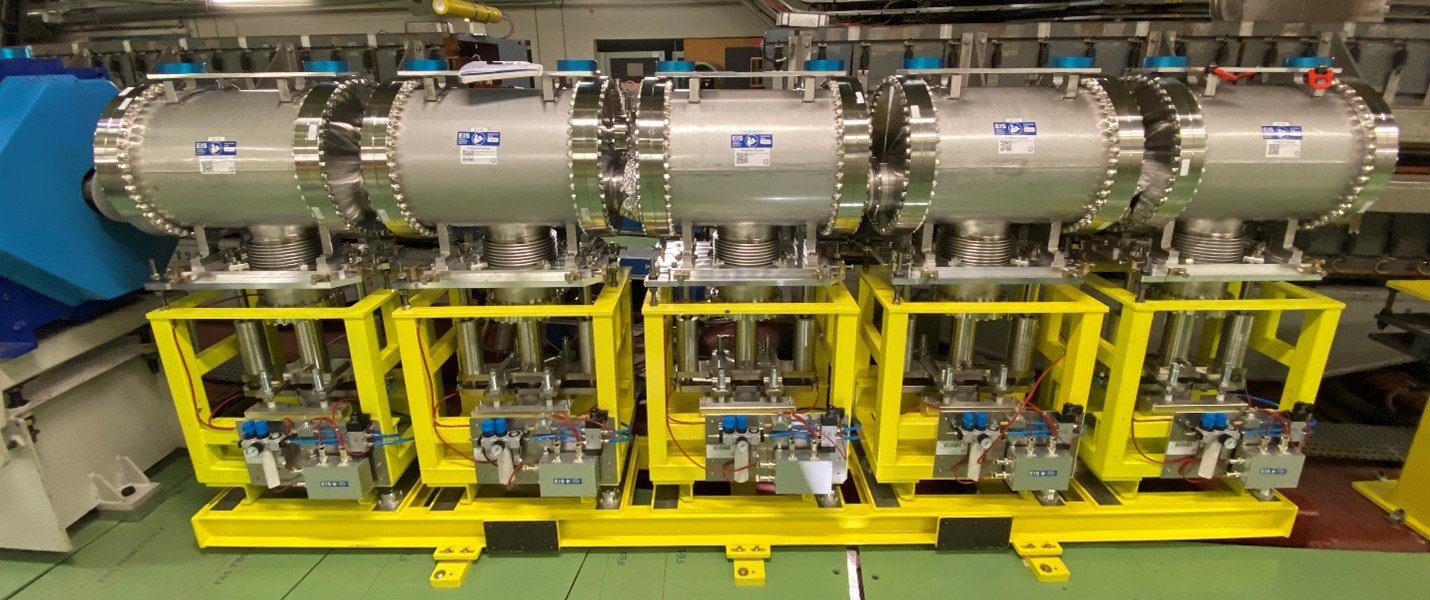}
\caption{Five beam stoppers installed on a girder.}
\label{Multiframe}
\end{figure}

For installation points requiring a high number of beam stoppers in a row, girders have been constructed to ease the installation and alignment processes (Fig.~\ref{Multiframe}).

\subsection{Operational feedback}
\begin{figure}[htb]
\centering
\includegraphics[width=0.48\textwidth]{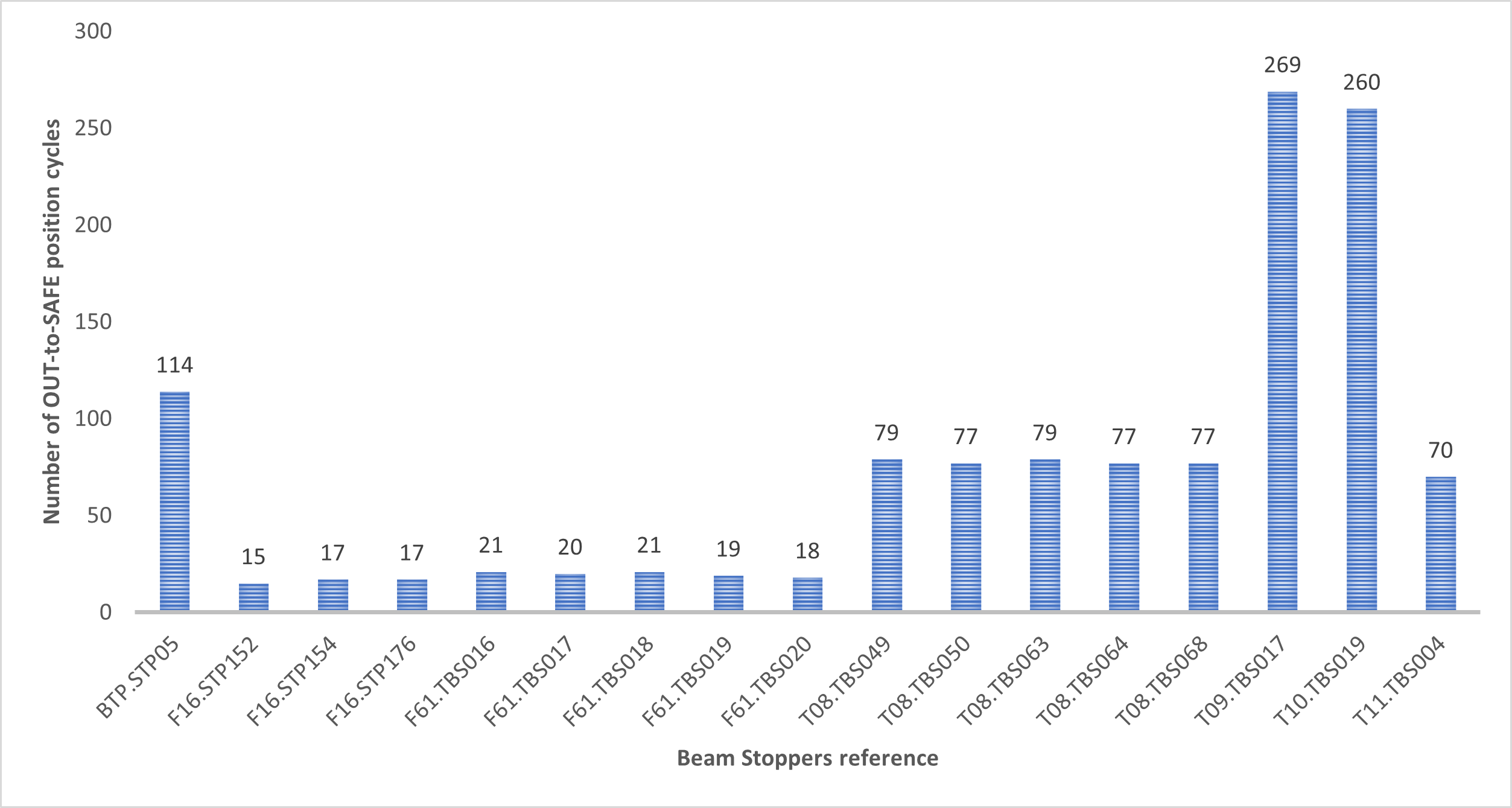}
    \caption{Numbers of out-to-in cycles for each of the 18 new beam stoppers installed from their installation date to the 2023 end-of-year technical stop, naming of the beam stoppers according to beam line and naming convention at CERN.}
    \label{graphcycles}
\end{figure}

To establish the total number of out-to-in-beam cycles, it is possible to crosscheck the upstream beam current transformer and beam-loss monitor data, or to directly extract the number of switch triggers from the CERN logging system for each installed beam stopper (Fig.~\ref{graphcycles}). Nevertheless, as noted earlier, this number of cycles does not represent the number of impacts; this could only be known by manually checking with the operation for incidents involving the in-beam positioning of a beam stopper.

Certain beam stoppers installed on the same line have differences in their numbers of cycles or only underwent a few cycles due to manual tests during preventative maintenance. The differences in term of cycles between the beam stoppers of the East Area facility---for example, T08 and T11 vs T09 and T10---is explained by different access frequencies due to operational requirements. Some beam stoppers are also used as stopper dumps, combining personnel protection and machine protection. By extrapolation, taking into account the cycles preformed over the first 3~years of the beam stoppers, the duty cycle expectations are stated as being between 150 and 2690 cycles after 30~years of operation, depending on the specific beam stopper, if the conditions of use remain constant.

It also has to be noted that the modular design increases flexibility in terms of the potential location of the beam stoppers in a given beam line. This allows the beam stoppers, especially those that were installed first, to be placed at locations that are better suited in terms of the radiation generated by the beam impacting them. Typically, this means that they can be placed in locations that are better shielded or are far from access points to the machine.

\section{Conclusion}
\label{Conclusion}
The obsolescence of the old beam stoppers in terms of stroke time, lifespan, and inappropriate cores for the LHC upgrade, as well as a lack of documentation and their several decades of operation, mean that they can no longer reliably ensure the safety of personnel.

The new beam stoppers can meet the requirements of the increase in beam-pulse energy to 92.5~kJ, replacing six different designs and offering the advantages of a unique, standard, but adaptable design that takes into account the beam parameters, attenuation, vacuum, and integration constraints. The mechanical performance of the new beam stoppers will enable better protection of personnel for the coming years by reducing the actuation time to 3~s and allowing the technical team to perform visual inspections and stopper-core changes.

Two aspects of the new beam-stopper design are significant for radiological safety: the modularity of the system and the material choices for the stopper core. The modularity of the system offers more flexibility in terms of choice of location within the beam line. Therefore, it is easier to choose a location that is well shielded, minimizing the radiological impact of the generated stray radiation. In addition, this modularity eases the realization of the total hadronic interaction length of a beam-stopper system required for a given beam line within its constraints such as beam optics or practicality of integration. The material used for the stopper core---Inconel~718 and CuCr1Zr---are a very good compromise between activation properties and the desire to have a short hadronic interaction length.

All these design decisions mean that whatever the installation area, the installation and removal processes will be identical and notably easier than for previous designs. Moreover, the increased ease and speed of installation and/or stopper-core replacement help to meet the ALARA principle.

\section*{Data availability}
Data will be made available on request.

\section*{Acknowledgments}
The authors would like to thank: the design office and the radiation-protection team for their involvement in the achievement of the project; the whole project team for their responsiveness and their support; and Calum Sharp as well as Edouard Grenier-Boley for proofreading. The authors would like to acknowledge the support of CERN's Sources, Targets, and Interactions (STI) Group in this project.

\section*{Declaration of competing interest}
The authors declare that they have no known competing financial interests or personal relationships that could have appeared to influence the work reported in this paper.

\bibliography{Bibliography}

\begin{thebibliography}{10}
\expandafter\ifx\csname url\endcsname\relax
  \def\url#1{\texttt{#1}}\fi
\expandafter\ifx\csname urlprefix\endcsname\relax\def\urlprefix{URL }\fi
\expandafter\ifx\csname href\endcsname\relax
  \def\href#1#2{#2} \def\path#1{#1}\fi

\bibitem{Maglioni}
C.~Maglioni, R.~Folch, Definition of beam stoppers, beam dumps and
  stopper/dumps, eDMS 1283556, CERN (2015).

\bibitem{Damerau:1976692}
J.~Coupard, H.~Damerau, A.~Funken, R.~Garoby, S.~Gilardoni, B.~Goddard,
  K.~Hanke, A.~Lombardi, D.~Manglunki, M.~Meddahi, B.~Mikulec, G.~Rumolo,
  E.~Shaposhnikova, M.~Vretenar (Eds.), {LHC} {I}njectors {U}pgrade, Technical
  Design Report, Vol. I: {P}rotons, 2014.
\newblock \href {https://doi.org/10.17181/CERN.7NHR.6HGC}
  {\path{doi:10.17181/CERN.7NHR.6HGC}}.

\bibitem{Pilan}
A.~P. Zanoni, J.~A.~B. Monago, E.~Grenier-Boley, M.~Calviani, V.~Vlachoudis,
  Characterization and core renovation of beam stoppers for personnel safety,
  J. Instrum. 14~(01) (2019) T01011.
\newblock \href {https://doi.org/10.1088/1748-0221/14/01/T01011}
  {\path{doi:10.1088/1748-0221/14/01/T01011}}.

\bibitem{ASTM}
Astm e8 standard, {https://www.astm.org/e0008-e0008m-22.html}.

\bibitem{EN12420}
En 12420 standard,
  {https://www.boutique.afnor.org/fr-fr/norme/nf-en-12420/cuivre-et-alliages-de-cuivre-pieces-forgees/fa150029/43577}.

\bibitem{SLAC}
S.~Rokni, A.~Fass{\`o}, J.~Liu, Operational radiation protection in high-energy
  physics accelerators, Radiat. Prot. Dosim. 137~(1-2) (2009) 3--17.
\newblock \href {https://doi.org/10.1093/rpd/ncp194}
  {\path{doi:10.1093/rpd/ncp194}}.

\bibitem{SPIRAL2}
E.~Schibler, J.-C. Ianigro, J.~Morales, N.~Redon,
  \href{https://accelconf.web.cern.ch/LINAC2010/papers/MOP097.pdf}{{Design of a
  high energy beam stop for Spiral2 }}, in: Proceedings of Linear Accelerator
  Conference LINAC2010, Vol. MOP097, Tsukuba, Japan, 2010, pp. 283--285.
\newline\urlprefix\url{https://accelconf.web.cern.ch/LINAC2010/papers/MOP097.pdf}

\bibitem{GARIS2}
S.~Kimura, D.~Kaji, Y.~Ito, H.~Miyatake, K.~Morimoto, P.~Schury, M.~Wada,
  Reduction of contaminants originating from primary beam by improving the beam
  stoppers in {GARIS-II}, Nucl. Instrum. Methods Phys. Res. A 992 (2021)
  164996.
\newblock \href {https://doi.org/10.1016/j.nima.2020.164996}
  {\path{doi:10.1016/j.nima.2020.164996}}.

\bibitem{Fermilab}
{F}ermilab {C}ontrolled {A}ccess {H}andout,
  \href{https://www-esh.fnal.gov/CourseHandout\_Mat/Cont\_Access/Cont\_Access.pdf}{https://www-esh.fnal.gov/CourseHandout\_Mat/Cont\_Access/Cont\_Access.pdf}
  (2005).

\bibitem{CEI}
Cei61508 standard,
  {https://www.boutique.afnor.org/fr-fr/norme/iec-6150812010/securite-fonctionnelle-des-systemes-electriques-electroniques-electroniques/xs121824/244493}.

\bibitem{FLUKA1}
C.~Ahdida, D.~Bozzato, D.~Calzolari, F.~Cerutti, N.~Charitonidis, A.~Cimmino,
  A.~Coronetti, G.~D’Alessandro, A.~Donadon~Servelle, L.~Esposito, et~al.,
  New capabilities of the {FLUKA} multi-purpose code, Front. Phys. 9 (2022)
  788253.
\newblock \href {https://doi.org/10.3389/fphy.2021.788253}
  {\path{doi:10.3389/fphy.2021.788253}}.

\bibitem{FLUKA2}
G.~Battistoni, T.~Boehlen, F.~Cerutti, P.~W. Chin, L.~S. Esposito,
  A.~Fass{\`o}, A.~Ferrari, A.~Lechner, A.~Empl, A.~Mairani, et~al., Overview
  of the {FLUKA} code, Ann. Nucl. Energy 82 (2015) 10--18.
\newblock \href {https://doi.org/10.1016/j.anucene.2014.11.007}
  {\path{doi:10.1016/j.anucene.2014.11.007}}.

\bibitem{FLUKA3}
{FLUKA} website, \href{https://fluka.cern}{https://fluka.cern}.

\end{thebibliography}
\bibliographystyle{elsarticle-num}

\end{document}